\documentclass[preprint,preprintnumbers, superscriptaddress, amsmath,amssymb]{revtex4}

\usepackage{txfonts}
\usepackage{hyperref}
\usepackage{graphicx}

\usepackage{setspace}
\usepackage{color}
\usepackage[sort&compress]{natbib}
\usepackage{bm}

\begin{document}

 \begin{center}
--Draft--
 \end{center}


\title{Surface waves in photonic crystal slabs}

\author{B. Wang}
\affiliation{%
Ames Laboratory and Department of Physics and Astronomy, Iowa State
University, Ames, Iowa 50011, USA
}%
\author{W. Dai}%
\affiliation{%
Ames Laboratory and Department of Physics and Astronomy, Iowa State
University, Ames, Iowa 50011, USA
}%
\author{A. Fang}%
\affiliation{%
Ames Laboratory and Department of Physics and Astronomy, Iowa State
University, Ames, Iowa 50011, USA
}%
\author{L. Zhang}%
\affiliation{%
Department of Electrical and Computer Engineering and
Microelectronics Research Center, Iowa State University, Ames, Iowa
50011, USA}

\author{G. Tuttle}%
\affiliation{%
Department of Electrical and Computer Engineering and
Microelectronics Research Center, Iowa State University, Ames, Iowa
50011, USA}

\author{Th. Koschny}%
\affiliation{%
Ames Laboratory and Department of Physics and Astronomy, Iowa State
University, Ames, Iowa 50011, USA
}%

\affiliation{%
Institute of Electronic Structure and Laser, FORTH, and Department
of Materials Science and Technology, University of Crete, Heraklion, Crete, Greece
}%

\author{C. M. Soukoulis}%
 \email{soukoulis@ameslab.gov}
\affiliation{%
Ames Laboratory and Department of Physics and Astronomy, Iowa State
University, Ames, Iowa 50011, USA
}%
\affiliation{%
Institute of Electronic Structure and Laser, FORTH, and Department
of Materials Science and Technology, University of Crete, Heraklion, Crete, Greece
}%

\begin{abstract}
Photonic crystals with a finite size can support surface modes when
appropriately terminated. We calculate the dispersion curves of
surface modes for different terminations using the plane wave
expansion method. These non-radiative surface modes can be excited
with the help of attenuated total reflection technique. We did
experiments and simulations to trace the surface band curve, both in
good agreement with the numerical calculations.
\end{abstract}


 \maketitle

The most well-known feature of Photonic crystals (PCs) is the
photonic band gap (PBG), inside which the electromagnetic waves are
prohibited to propagate in all directions \cite{book Soukoulis}. However, when
appropriately terminated, PCs can support surface modes with
frequencies lying inside the PBG \cite{book PC}.  Surface modes are easily obtained for one-dimensional PCs (stacks of alternating continuous layers) and their dispersion relation has been obtained experimentally and theoretically\cite{sw3}.

For a 2D PC the existence of surface waves has been shown theoretically (in band structure calculations for semi-infinite PC) \cite{sw1, sw2} and experimentally by attenuated total reflection (ATR) measurements \cite{mit atr}. Further numerical work has shown that the existence and dispersion of surface waves at a 2D PC are very sensitive to the surface termination \cite{sw2, observe surface state}. When terminated
improperly, both the poor impedance matching and the interaction with
surface modes can  disrupt the electromagnetic wave propagation and
are bad to the performance of photonic devices such as PC waveguides
and PC cavities \cite{observe surface state}. Recent studies also
show that when the PC is appropriately periodically corrugated  at
the surface, enhanced transmission and beaming of light can be
achieved through a PC waveguide with a subwavelength width
\cite{beaming 1, beaming 2, ozbay} . This effect is due to the
excitation of surface modes and constructive interference at the
axis of the waveguide.

In this paper, we study experimentally and numerically the dispersion relation of surface modes for different surface corrugations that support surface waves. In contrast to previous work, we determine experimentally the complete dispersion relation of the surface modes within the photonic band gap of the PC.

In addition to the well-known theoretical band-structure calculations for the infinite periodic PC we also perform full-wave frequency-domain finite element (FEMLAB) simulations to determine the surface mode dispersion relation numerically for a finite PC and compare the three different approaches.

\begin{figure}
\begin{center}
  \includegraphics[width=8cm]{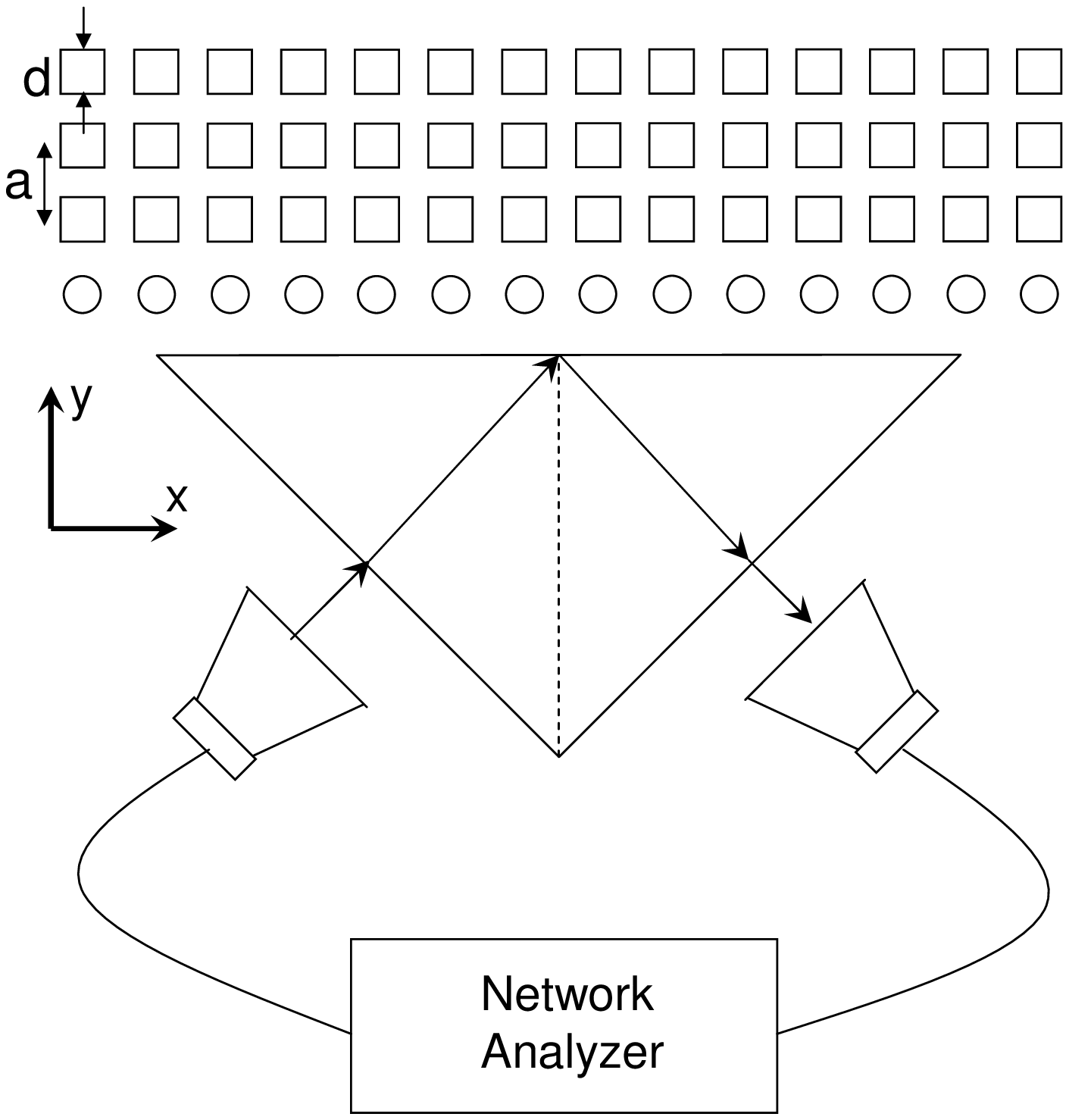}\\
  \end{center}
  \caption{Experimental setup. The HP8510 network analyzer and
  a pair of horn antennas are used to measure the reflection spectrum of the prism.
The PC, with lattice constant $a=11 \, mm$ and dimension of the  square
rod $d=3.1 \, mm$, is put behind the prism.}\label{Fig:setup}
\end{figure}

\begin{figure}
\begin{center}
  \includegraphics[width=14cm]{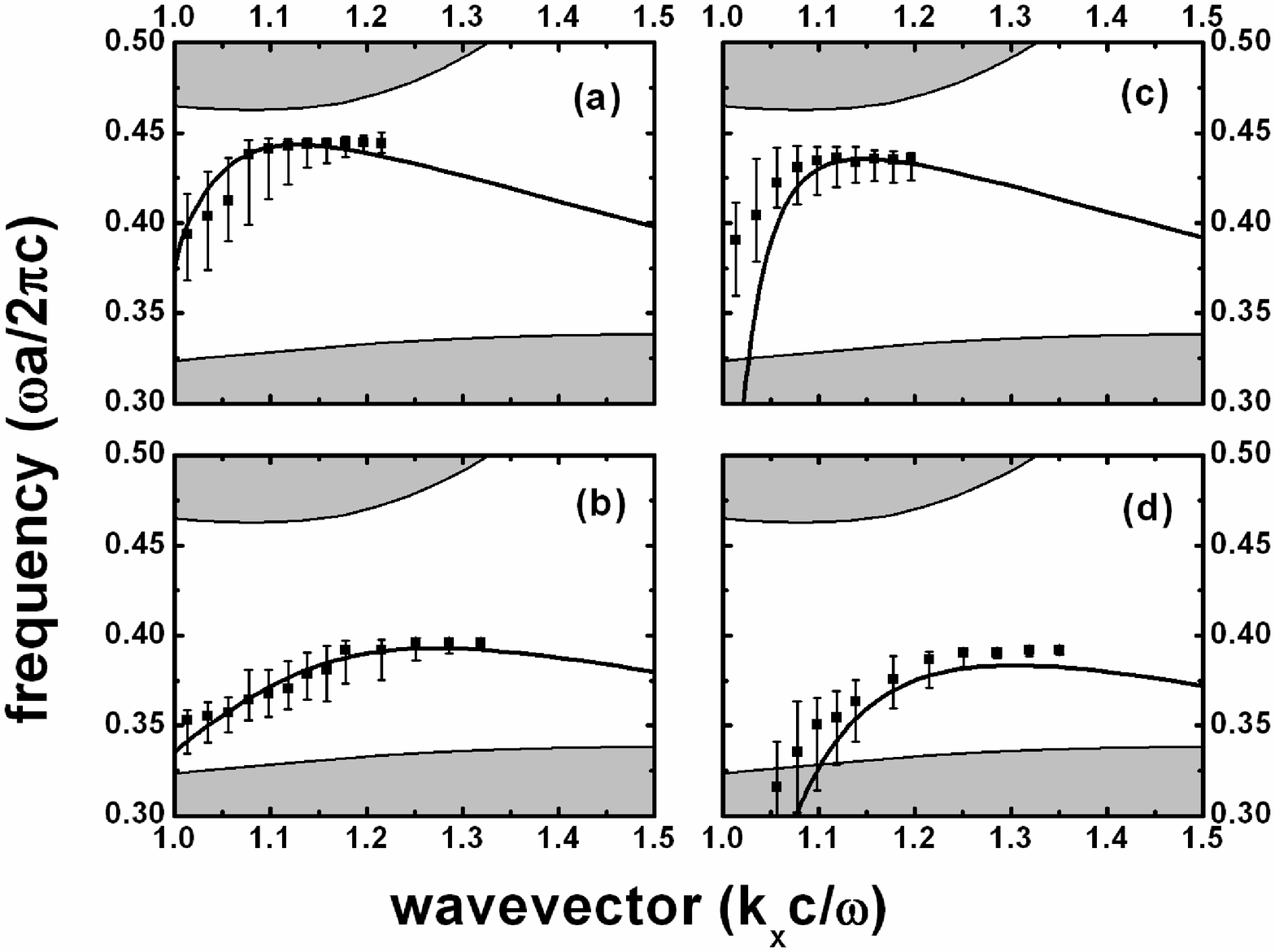}\\
  \end{center}
  \caption{Experimental surface band structure of a 2D PC.
Between the light gray areas exist the band gap of PC. Solid lines
and square dots with error bars are supercell band structure
calculation and experimental results of the surface band for (a) PC
with a corrugation layer of circular rods with $D=1.83 \, mm$, (b)
PC with a corrugation layer of circular rods with $D=2.44 \, mm$,
(c) a single layer of rods with $D=1.83 \, mm$ and (d) a single
layer of rods with $D=2.44 \, mm$.}\label{Fig:ATR_exp}
\end{figure}

The 2D PC we study is a square array of
square alumina  rods, 21 layers in $x$ direction and 15 layers in
$y$ direction. The lattice constant is $a=11 \, mm$ and the square
rods are of dimension $d=3.1  \, mm$ with permittivity $9.8$ and
height $h=15 \, cm$ (Fig. \ref{Fig:setup}). With the help of the
supercell technique, the band structure of the finite-size PC can
be calculated using the plane wave expansion method \cite{1990 Ho}.
We studied TM (electric field along the rods) modes through out the
work since only TM modes can give a full band gap for a two
dimensional PC \cite{book PC}. As shown, for example, in
Fig.\ref{Fig:ATR_exp} (a), the PC has a band gap in between the
light gray areas. There are no modes inside the band gap.

The uncorrugated PC does not support surface waves (in the lowest total band gap); to observe surface modes we have to corrugate the surface layer of the PC, rendering it different from the air on one side and the PC bulk on the other side.

The actual shape of the corrugation layer elements is not very important. We use cylindrical rods of different diameters, but also cut semi-cylinders \cite{mit atr}, rectangular cross-section rods different from the ones comprising the bulk PC \cite{square corrugation} or geometrically identical rods with different dielectric constant might be used. However, there is no guarantee to obtain surface modes for a particular surface termination; the surface layer parameters have to be chosen carefully for the corrugated PC to support surface modes in the band gap of the bulk PC.

In order to support surface modes, we add  a new layer at the
surface of the PC with circular alumina rods as a periodic
corrugation (Fig. \ref{Fig:setup}). The band structure of the
corrugated PC was calculated using the plane wave expansion and the
supercell technique. The solid lines in Figures \ref{Fig:ATR_exp}(a)
and \ref{Fig:ATR_exp}(b) show the dispersion relation of the surface
modes which lie  inside the band gap when the diameter of the
circular rods is $D=1.83  \, mm$ and $D=2.44   \,mm$, respectively.
A surface mode cannot extend into the PC since the frequency is
within the band gap. Nor can it extend into air because the surface
mode lies to the right of the light line ($k_{x}c/\omegaup>1$). So
this mode is localized to the surface of the structure and can be
called a surface mode. The diameter of the corrugation rods needs to
be chosen properly so that the surface dispersion lies within the
band gap. When the surface mode goes out of the band gap, the mode
can extend into the PC and is no longer a surface mode. Actually, a
single layer of only dielectric rods surrounded by air can support
surface modes. The solid line in Fig. \ref{Fig:ATR_exp} (c) and (d)
shows the surface band of a single layer of circular alumina rods
with diameter $D=1.83  \,mm$ and $D=2.44   \,mm$, respectively.
These two surface dispersion curves shown in Figures
\ref{Fig:ATR_exp}(c) and \ref{Fig:ATR_exp}(d)  are different from
the two (shown in Fig. \ref{Fig:ATR_exp} (a) and (b)) which also
involve the PC. This can be explained from the fact that the
presence of the PC modifies the surface mode of a single layer.

Since surface modes lies to the right of the light line, they cannot
be excited by propagating waves directly. To observe a surface mode
experimentally, we use the ATR method
(Otto configuration) \cite{book SP}. Consider a dielectric prism
with index of refraction $n$. If the incident angle $\theta$ at the
reflecting surface is larger than the critical angle $\theta_{C}$,
$\theta>\theta_{C}=\sin^{-1}(1/n) $, total internal reflection
occurs at the outgoing interface. When there is nothing behind, the
incident wave would be totally reflected. However, apart from the
reflected wave, total reflection also involves a transmitted
evanescent wave which travels along the interface and decays
exponentially away from the surface. That is because the wavevector
component parallel to the interface,
\begin{equation}\label{eq:k//}
    k_{x}=k\sin{\theta}=n\omegaup\sin{\theta}/c>\omegaup/c,
\end{equation}
is conserved across the interface. So outside the prism, the
wavevector component perpendicular to surface,  $k_{y}$, is
imaginary. When we put the corrugated PC close to the prism, the
tail of this evanescent wave may couple to a surface mode at the
surface of the PC. By changing the incident angle, we can change the
value of $k_{x}$ and get coupling with different surface modes.  By
plotting the experimental results at different incident angles, the
surface band can be obtained.

The first ATR experiment on surface modes of PC was reported  more
than 10 years ago \cite{mit atr}. However, it was not
straightforward to see the coupling of surface waves from the
experimental data and a dispersion relation was not obtained
experimentally. In our experiment, an HP-8510 network analyzer was
used to measure the S parameters so that the reflection spectrum can
be immediately seen. A pair of horn antennas serves as transmitter
and receiver (Fig. \ref{Fig:setup}). The dielectric prism used in
our experiments is an isosceles-right-triangle-shaped wedge with
side length of $15 \,cm$ and with index of refraction $n=1.61$.  The
horn antennas were placed in such a way that the E field is
polarized along the dielectric rods. For a given incident angle,
Snell's law was used to identify the angle and position of the horn
antennas such that the setup is symmetric along the dashed line.

\begin{figure}
\begin{center}
  \includegraphics[width=10cm]{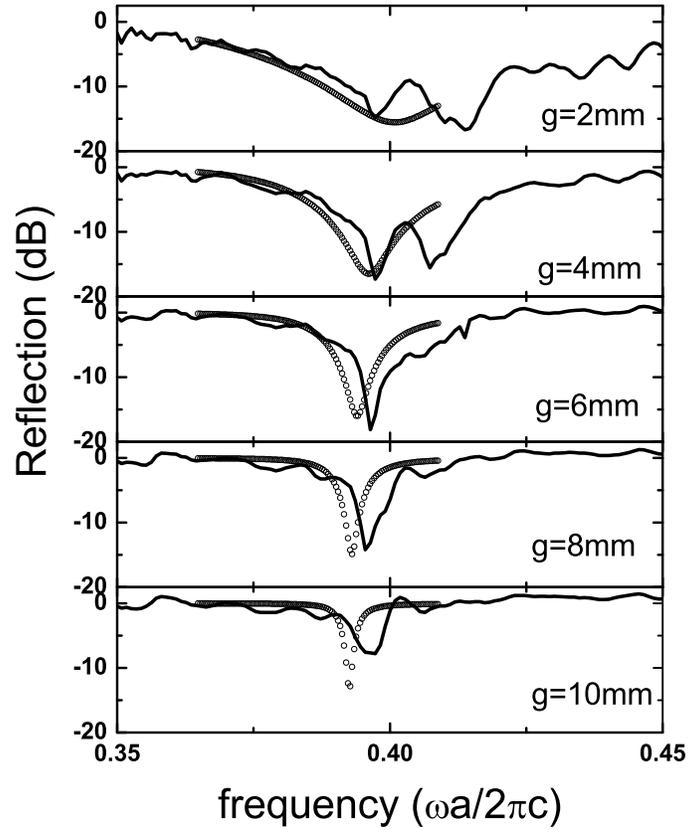}\\
   \end{center}
\caption{The experiment (solid lines) and simulation (circles)
results of reflection at incident angle $53\textordmasculine$ for
the PC with corrugation layer of circular rods with $D=2.44 \, mm$.
From top to bottom shows the reflection when  the distance $g$ of
the air gap between the prism and the surface of the structure
increases.}\label{Fig:53}
\end{figure}

\begin{figure}
\begin{center}
\includegraphics[width=10cm]{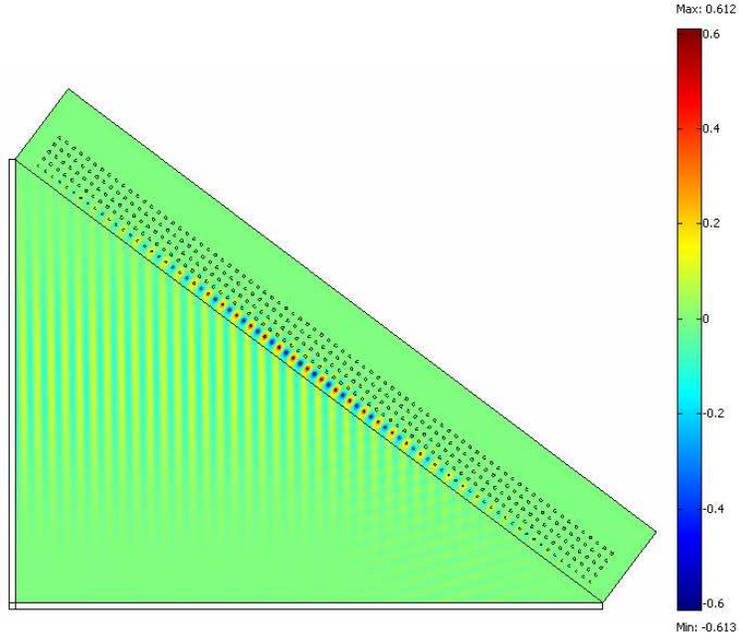}\\
\end{center}
\caption{(Color online) $E_{z}$ (E field component along the rods) distribution
when a surface mode is excited by simulation. The incoming wave is
normally incident from the left of the prism so that the incident
angle is $53\textordmasculine$ at the reflecting surface. Behind the
prism is four layers of PC rods with one corrugation layer of
circular rods with diameter $D=2.44 \, mm$. The airgap size between
the prism and the structure is $g=10\, mm$. The normalized frequency
of the incident wave is $0.39$. In the figure black and white areas
have strong field and light gray areas have relatively small
field.}\label{Fig:freq107}
\end{figure}

The ATR method requires the structure to be set close to the prism
to get good coupling effect because the evanescent waves decay fast;
however, while we are making use of the prism, the prism itself
modifies the surface modes especially when it is very close to the
structure. To get a satisfying experimental  result, we need to
optimize the airgap size to be able to record the mode while keeping
the structure as far away from the prism as possible. In Fig.
\ref{Fig:53} the solid curves show the coupling effect at different
air gap sizes at incident angle $53\textordmasculine$ for the PC
with corrugation layer of circular rods with diameter $D=2.44 \,
mm$. You can see clearly in Fig. \ref{Fig:53} that when the surface
waves are excited, there is a very well defined dip in the
reflection data. This is well shown for the distance of $g=10 \,
mm$. In Fig. \ref{Fig:freq107}, one sees clearly that a surface wave
is excited and almost no reflection is seen. Fig. \ref{Fig:freq107}
is an illustration of field distribution when a surface mode is
excited, given by simulation. A surface wave with intensity much
stronger than the incoming wave is seen along the corrugation layer of
the PC. Actually the intensity of the surface wave is so strong that
the reflected wave is hardly seen. This effect is observed as a dip
in the reflection spectrum of the prism in our experiments (see Fig.
\ref{Fig:53} for $g=10 \, mm$). The prism disturbs the coupling and
the observed reflection dip position when the air gap is small; the
dip converges to a constant frequency as the air gap increases. At
larger incident angles, the coupling  of the evanescent waves to
surface waves is smaller and the dip position converges faster. This
is because larger incident angle provides larger $k_{x}$ and thus
larger absolute value of imaginary $k_{y}$; so the evanescent wave
decays faster. For the surface corrugated PC, the ATR dip is
still strong enough to be observed when the air gap between the
prism and the surface of the structure is $10 \, mm$ and is already
converged except at incident angles near the critical angle. For a
single layer of corrugation rods, the coupling is not as strong as
for the full structure (PC and corrugation layer) case. To get good
coupling, for the single corrugation layer, the air gap size is
chosen to be $7 \, mm$.

Fig.\ref{Fig:ATR_exp} shows the experimental results of the  surface
bands together with the calculation results for two different
corrugation layers with and without PC. The solid points are the
averaged dip frequencies over three measurements while the error bar
on each point comes from the half width of the reflection dip,
averaged over three different measurements. We can see that the
experimental results are in good agreement with the calculated
surface bands. However, at smaller incident angles, especially close
to the critical angle, relatively large error bars are obtained.
This is because the finite-width incident beam might not be fully
reflected; some plane wave components propagate through the
interface and change the coupling to the surface mode. Instead of a
nice and sharp coupling dip, the coupling now is over a larger
frequency range. That makes the half width of the dip larger. Also
the evanescent wave decays slower at smaller $k_{x}$ and the
coupling may not have converged yet at the largest air gap. So the
overall experimental error is larger.

\begin{figure}
\begin{center}
  \includegraphics[width=14cm]{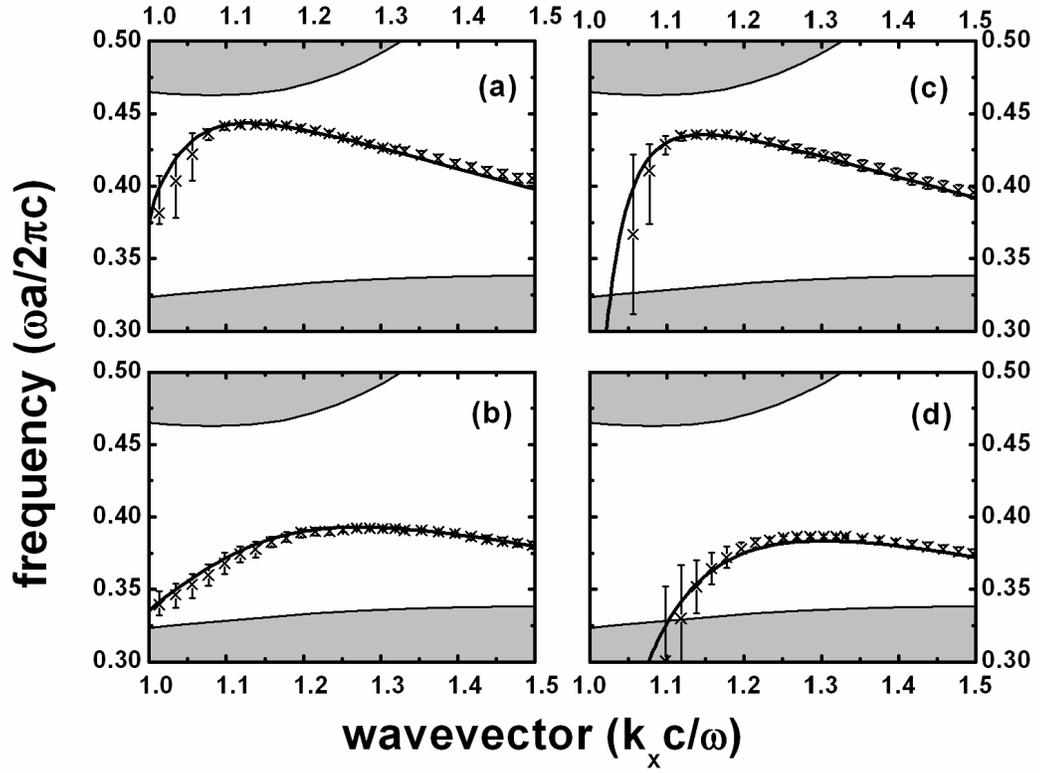}\\
   \end{center}
 \caption{Simulated surface band structure of a 2D PC.
Between the light gray areas exist the band gap of PC. Solid lines
and crosses with error bars are supercell band structure calculation
and simulation results of the surface band for (a) PC with a
corrugation layer of circular rods with $D=1.83 \, mm$, (b) PC with
a corrugation layer of circular rods with $D=2.44 \, mm$, (c) a
single layer of rods with $D=1.83 \, mm$ and (d) a single layer of
rods with $D=2.44 \, mm$.}\label{Fig:ATR_sim}
\end{figure}

Simulations with a similar setup were done in FEMLAB and the results
are shown in Fig.\ref{Fig:ATR_sim}.  In the simulations, the source
is a Gaussian beam normally incident at one side of a right-angled
dielectric wedge. The rod array is set behind the hypotenuse of the
wedge. The reflected power is obtained by integrating over the other
side of the wedge. To get a coupling at different angles, the two
acute angles of the wedge are changed so that the Gaussian beam is
always normally incident on the wedge. Compared with experiments
(Fig. \ref{Fig:53}), the reflection spectrum in simulations is always
smoother. Also, the coupling can still be seen when the air gap is
several centimeters large. The simulation results are in excellent
agreement with the supercell calculations, as shown in Fig. \ref{Fig:ATR_sim}.

In conclusion,we have investigated the excitation of surface waves on surface-corrugated PCs as well as single PC layers. ATR experiments and simulations have been done to
excite different surface modes and the dispersion relations have been
obtained, both being in good agreement with numerical calculations
using the plane wave expansion method and the supercell method.

We gratefully acknowledge the support of Ames Laboratory, which is
operated by Iowa State University under contract No. W-7405-Eng-82,
EU projects PHOREMOST, METAMORAHOSE and DARPA contract No.
HR0011-05-C-0068. This work was supported by the AFOSR under MURI grant FA9550-06-1-0337.


\newpage

\end{document}